\documentclass{IEEEtran4PSCC}
\usepackage[pdftex]{graphicx}

\usepackage{siunitx}
\usepackage[cmex10]{amsmath}
\usepackage{makecell}

\makeatletter
\let\old@ps@headings\ps@headings
\let\old@ps@IEEEtitlepagestyle\ps@IEEEtitlepagestyle
\makeatother

\begin{document}
\title{A Physics Informed Machine Learning Method for Power System Model Parameter Optimization}

\author{
\IEEEauthorblockN{Georg Kordowich\\ Johann Jaeger}
\IEEEauthorblockA{Institute of Electrical Energy Systems \\
Friedrich-Alexander-Universität Erlangen-Nürnberg\\
Erlangen, Germany\\
georg.kordowich@fau.de}
}

\maketitle
\texttt{}

\begin{abstract}
This paper proposes a gradient descent based optimization method that relies on automatic differentiation for the computation of gradients. The method uses tools and techniques originally developed in the field of artificial neural networks and applies them to power system simulations. It can be used as a one-shot physics informed machine learning approach for the identification of uncertain power system simulation parameters. Additionally, it can optimize parameters with respect to a desired system behavior. The paper focuses on presenting the theoretical background and showing exemplary use-cases for both parameter identification and optimization using a single machine infinite busbar system. The results imply a generic applicability for a wide range of problems.
\end{abstract}


\section{Introduction}
\subsection{Background}
The growth in distributed generation increases the complexity of the grid. More variability in generation and loadflow demands reinforcements for the power grid. As the construction of new powerlines and grid components can be prohibitively expensive, many recent approaches focus on maximizing the utilization of existing infrastructure. Consequently, a trend towards a smarter grid is notable, characterized by increased automation of grid operations and the installation of sensors and actors to enhance grid observability and controllability.

Accurate models of power systems and components are an essential foundation of the trend towards a smarter grid. The utilization of concepts like digital twins is becoming increasingly prevalent for grid control and supervision \cite{wagner}. For concepts like model predictive control, an accurate mathematical description is a key prerequisite \cite{Mahr.2021b}. Additionally, a lot of recent research focuses on employing different optimization or machine learning techniques in order to improve grid stability and security. Examples are the optimization of protection schemes, energy management, demand response or operational control \cite{Kordowich, .2019}.

\subsection{Challenge}
While good models are a fundamental requirement for all aforementioned advancements, it is often time consuming and difficult to create accurate models. Even though the general structure of a model or digital twin may be known, the task of finding precise parameters that match real world behavior is often particularly difficult.

Therefore, parameter identification and optimization can be identified as a core challenge for enabling secure operation in future power systems. While often described as two different tasks, it is important to note that parameter identification and optimization are essentially the same problem from a mathematical point of view: Parameter identification is in fact an optimization problem, aiming to minimize the error between model and real-world behavior. 

\subsection{Automatic Differentiation in Power Systems}
To address the challenge of parameter optimization and identification in power system simulations we propose a novel method that incorporates an automatic differentiation (AD) tool into a dynamic power system simulation. AD tools can be used to compute gradients of mathematical functions. In the context of our method, those gradients are utilized to optimize simulation parameters by minimizing a loss or error function using gradient descent.

The idea of using AD tools in the context of power systems was previously employed for a range of applications. In the steady state domain, they were used to calculate the Jacobian matrix for power flow calculations \cite{powerflow1}, to model power electronics \cite{powerflow3}, or for state estimation \cite{steadystateest}. In the dynamic domain, AD tools have been used to create an induction machine model in the resistive companion form \cite{Gao.2004} and for trajectory sensitivity analysis \cite{Geng.2014}.

Another method in the dynamic domain that utilizes AD tools and are related to our approach are Physics Informed Neural Networks (PINNs) \cite{Misyris.09Nov19}. The core idea behind PINNs is the embedding of analytical models into neural networks which benefits training speed and reduces the maximum training error bounds \cite{Maier.2019}. PINNs have previously been used for a wide range of tasks \cite{Huang.2023}. Similarly to our approach, PINNs are also capable of system parameter identification~\cite{Stiasny.2021b}.

\begin{figure}
	\centering
	\includegraphics[width=1\linewidth]{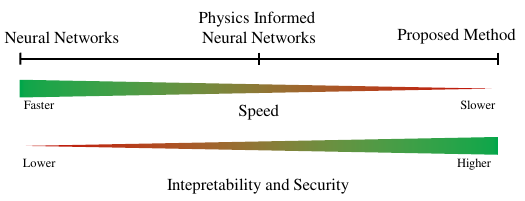}
	\caption{Categorization of different Machine Learning Methods for Power System Simulation}
	\label{fig:fasterslower}
\end{figure}

A drawback of PINNs is however, that physical knowledge covers only part of the learning process. Other parts that are either neglected or unknown, must be learned by a neural network that is essentially a black-box model. Consequently, a training process that requires data is necessary. Additionally, the use of neural networks can lead to instable or unforseeable behavior outside of the training data range \cite{Misyris.09Nov19}.

Our approach utilizes the same idea of incorporating physics knowledge into the optimization problem, but completely eliminates the neural networks and solely relies on the physical knowledge. Compared to neural networks (blackbox models) and PINNs (greybox models), our approach ensures maximum interpretability and security by only relying on equations with physical meaning. Therefore, the novel approach can be considered a whitebox model. This comes with a computational cost as shown in Fig. \ref{fig:fasterslower}. On the other hand, its advantages lie in not relying on training data and its capacity to identify parameters from a singular time series, rendering the approach capable of one-shot, physics informed machine learning.

Another physics informed machine learning approach that does not incorporate neural networks and can be used for parameter estimation is the sparse identification of nonlinear dynamics (SINDy) algorithm \cite{Lakshminarayana.2022}. It uses sparsity promoting algorithms to select equations from a set of candidate functions that fit the dynamic behavior of the system \cite{Hamid.2022}. SINDy requires not only time series data of state variables but also the derivatives. Another drawback is, that a reformulation of the power system description into matrix form is necessary, which limits the usability and generalizability of the approach.  

Professional power system simulation softwares often contain similar gradient descent based parameter optimization functions. To our best knowledge, all gradient estimation techniques used so far are based on the difference quotient in (\ref{eq:difq}), which makes small variations of the parameters to be optimized necessary.
\begin{equation}
	\frac{\partial f(x)}{\partial x} \approx \frac{f(x+h) - f(x)}{h}
	\label{eq:difq}
\end{equation}
This means that the number of simulations required per optimization step is equal to the number of parameters to be optimized. Using an AD tool, only one simulation is necessary for any number of parameters, making the presented approach computationally cheaper.
\newpage
The core contributions of this paper lie in:
\begin{itemize}
	\item introducing a physics informed optimization method for parameters of dynamic simulations;
	\item showing, that the optimization process is generically usable for a wide range of problems;
	\item and the publication of corresponding code for easy reproducibility \cite{Kordowich.2023}.
\end{itemize}

\section{Methodology}
In the following chapter, the optimization method is presented and explained in detail. For this purpose, first the general process behind power system simulations is explained in subsection \ref{sec:sim}. Afterwards, a brief introduction to optimization via gradient descent is given. Then, we show how simulation parameters can be optimized via gradient descent theoretically, and afterwards we show how AD tools make the theoretical approach practically viable. 

The methodology is shown using the example of the single machine infinite busbar (SMIB) system by Kundur \cite{Kundur.2007}. For the example, we assume the inertia constant $H$ of the generator is unknown. The goal of the process is the parameter identification of $H$, so that the dynamic response of the simulated generator matches the response of the original Kundur SMIB system. In practice, the same approach can be used to identify parameters of real world systems.

\subsection{Dynamic Power System Simulation}
\label{sec:sim}
An overview over power system simulations is given in the following subsection as a basis for the proposed method. For a more detailed descriptions interested readers are referred to \cite{Haugdal.08Jan21} or \cite{JanMachowskiJanuszW.BialekandJamesR.Bumby.}. Generally, dynamic power system models can be described by a set of Differential Algebraic Equations (DAE) shown in (\ref{eq:daeset}):
\begin{equation}
	\begin{aligned}
		\mathbf{\dot{x}} &= f(\mathbf{x}, \mathbf{y}) \\
		0 &= g(\mathbf{x}, \mathbf{y})
	\end{aligned}
	\label{eq:daeset}
\end{equation}
Here, $\mathbf{x}$ represents the vector of state variables while $\mathbf{y}$ contains the algebraic variables. The equation above can further be simplified by describing the algebraic variables $y$ as a function $h$ of differential variables $x$:
\begin{equation}
	\mathbf{\dot{x}} = f(\mathbf{x}, h(\mathbf{x})) = f_{new}(\mathbf{x})
	\label{eq:daesimp}
\end{equation}

In the case of the SMIB system, the differential equations represent the 6th order model of a generator where the state variables can be found on the left hand side \cite{JanMachowskiJanuszW.BialekandJamesR.Bumby.}:
\begin{equation}
	\begin{aligned}
		\Delta \dot{\omega} &= \frac{1}{2H} (T_m - T_e) - D\Delta\omega \\
		\dot{\delta} &= \Delta \omega \\
		\dot{E_q}' &= \frac{1}{T_{d0}'} (E_f - E_q' - (X_d - X_d') I_d) \\
		\dot{E_d}' &= \frac{1}{T_{q0}'} (-E_d' + (X_q - X_q') I_q) \\
		\dot{E_q}'' &= \frac{1}{T_{d0}''} (E_q' - E_q'' - (X_d' - X_d'') I_d') \\
		\dot{E_d}'' &= \frac{1}{T_{q0}''} (E_d' -E_d'' + (X_q' - X_q'') I_q')
	\end{aligned}
\end{equation}

The algebraic variables correspond to the bus-voltages, which can be calculated using the admittance matrix and the current injections. As the current injections are a function of the state variables, the algebraic variables can be represented as a function $h(x)$ of the state variables:
\begin{equation}
	\mathbf{y} = \mathbf{V} = \mathbf{Y}^{-1} \mathbf{I}(\mathbf{x}) = h(\mathbf{x})
\end{equation}

Together, the equations above fully describe the power system model. During a simulation, (\ref{eq:daesimp}) can be integrated by using suitable integration methods like Runge-Kutta or Euler's method. When using Euler's method, the trajectories of the state variables can be determined by applying the equation below during every timestep: 
\begin{equation}
	\mathbf{x_{t+1}} = \mathbf{x_{t}} + \mathbf{\dot{x}_t} \Delta t = \mathbf{x_{t}} + f(\mathbf{x_{t}}) \Delta t
\end{equation}

Therefore, a power system simulation consists of nothing but a long chain of very basic and most importantly locally differentiable operations. This fact is key for the applicability of the presented approach.

\subsection{Gradient Descent Optimization}
The gradient descent optimization method is a very common and well researched one. Its purpose is to optimize parameters in a way so that a \textit{loss} or an \textit{error} function, often a difference between desired and optimal output, gets minimized. This is achieved by adapting the parameters in the opposite direction of the gradient of the loss function with respect to the parameters. Intuitively, this can be described as "going the loss function downhill" towards a local minimum. Formally, the \textit{loss} or \textit{error} is often described by a function $L(\theta)$, that depends on the optimizable parameters $\theta$. The parameter adaption can formally be described by (\ref{eq:graddesc}):
\begin{equation}
	\mathbf{\theta}_{new} = \mathbf{\theta}_{old} - \eta \cdot \nabla_{\mathbf{\theta}} L(\mathbf{\theta})
	\label{eq:graddesc}
\end{equation}

The parameter $\eta$ is often referred to as the learning rate and describes how fast the optimizable parameters are changed. It can be adapted to balance the optimization process between fastness of convergence and stability.

In the example of the optimization of the SMIB system, an adequate loss function could be the mean squared error (MSE) between simulated rotor frequency and real rotor frequency, depending on the inertia constant $H_{sim}$:
\begin{equation}
		L(H) = \frac{1}{N}\sum_{t=1}^N(\Delta\omega_{real, t} - \Delta\omega_{sim, t}(H_{sim}))^2
	\label{eq:loss}
\end{equation}
Here, $N$ is the number of discretely measured or simulated timesteps for the real world and simulated trajectory of $\Delta\omega$ respectively. The gradient descent can then be executed by calculating the gradient of the loss function $L$ with respect to $H$. The Loss can be minimized, by iteratively adapting $H$ in the opposite direction of the gradients on $H$:
\begin{equation}
		H_{new} = H_{old} - \eta \frac{\partial L}{\partial H}
	\label{eq:graddescent}
\end{equation}

The same process works for multiple parameters at the same time by calculating the gradient $\nabla_\theta L(\theta)$.

The prerequisite of the gradient descent algorithm is of course, that the gradients on the parameters that must be optimized are known. To determine those gradients, the proposed method makes use of AD tools originally developed for neural networks. Those tools calculate the gradients of the loss function with respect to the weights of a neural network by iteratively applying the chain rule of differentiation. This process is known as backpropagation. The following section \ref{sec:bp} shows that the same process can be applied to power system simulations as well.

\subsection{Backpropagation for Power Systems}
\label{sec:bp}
Obtaining the gradient of the loss function with respect to $H$ or other parameters is a challenge. (\ref{eq:loss}) is a long and convoluted equation, as it depends on $\Delta\omega_{gen,sim}$ which depends on $H$ in a non-trivial way. On the other hand, the same is true for neural networks, where the loss function typically depends on thousands of parameters which influence the loss function in different ways. However, both neural networks, and power system simulations consist of a long chain of basic, and most importantly locally differentiable operations ($+$, $-$, $*$, $/$, $ln(x)$, $e^x$...) as shown in section \ref{sec:sim}.

This fact allows the application of the chainrule of differentiation. The gradient of the loss function with respect to the parameters does not have to be computed at once with an explicit expression, but instead it can be split up into small steps of gradients of intermediate basic operations:

\begin{equation}
	\frac{\partial L}{\partial H} = \frac{\partial L}{\partial o_1}\frac{\partial o_1}{\partial o_2}\frac{\partial o_2}{\partial o_3} \;\;  ... \;\; \frac{\partial o_{n-1}}{\partial o_{n}}\frac{\partial o_n}{\partial H}
	\label{eq:chainrule}
\end{equation}

For this to be feasible, it's essential that every operation performed during the simulation is recorded, enabling subsequent calculation of the local gradient. For this purpose AD tools like PyTorch or TensorFlow originally developed for the training of neural networks can be used \cite{Paszke.2017}. When training neural networks, during the forward pass (i.e. the evaluation phase of the neural network), those tools build a computational graph by recording every operation. When wanting to compute the gradients, this graph can be traversed backwards, by applying the chainrule of differentiation in every step. The same process can be applied for arbitrary functions and, as we show here, to power system simulations.

As an example suppose parameter $\theta$ shall be optimized so the (nonsensical) loss function given in (\ref{eq:nonsense}) gets minimized:
\begin{equation}
	L(\theta) = \left(a - \frac{b}{c \cdot \exp(\theta)}\right) - 0
	\label{eq:nonsense}
\end{equation}

\begin{figure}
	\centering
	\includegraphics[width=1\linewidth]{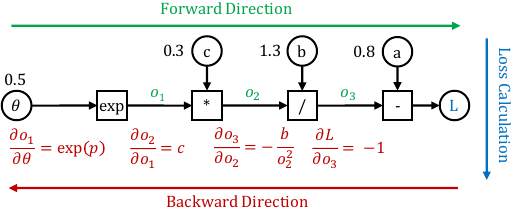}
	\caption{Computational graph for equation (\ref{eq:nonsense})}
	\label{fig:compgraph}
\end{figure}
\begin{table}[h!]
	\centering
	\renewcommand{\arraystretch}{1.4}
	\caption{Forward Direction}
	\label{table_values}
	\begin{tabular}{|c|c|c|c|c|}
		\hline
		$\theta$& $o_1$ & $o_2$ & $o_3$ & $L$\\
		\hline
		$0.5$ & $1.65$ & $0.49$ & $2.63$ & $-1.83$\\
		\hline
		\hline
		$\frac{\partial L}{\partial \theta}$ & $\frac{\partial o_1}{\partial \theta}$ & $\frac{\partial o_2}{\partial o_1}$ & $\frac{\partial o_3}{\partial o_2}$ & $\frac{\partial L}{\partial o_3}$ \\
		\hline
		$1.65$ & $0.30$ & $-5.41$ & $-1$ & $2.63$\\
		\hline
	\end{tabular}
\end{table}
Then, during the evaluation of the function, the AD tool would build the graph shown in Fig. \ref{fig:compgraph}, with the corresponding values for the (similarly nonsensical) exemplary values in the figure given in Table \ref{table_values}. Afterwards, the "local gradient" of every operation can be calculated in the backward direction as the derivatives of the basic operations are known. After the local gradients are calculated the gradient $\partial L / \partial \theta$ can directly be computed by applying (\ref{eq:chainrule}):
\begin{equation}
	\begin{aligned}
	\frac{\partial L}{\partial \theta} = &\frac{\partial L}{\partial o_3}\frac{\partial o_3}{\partial o_2}\frac{\partial o_2}{\partial o_1}\frac{\partial o_1}{\partial \theta} =\\ &1.65 * 0.30 * -5.41 * -1 = 2.63
    \end{aligned}
	\label{eq:chainruleapplied}
\end{equation}

\begin{figure}
	\centering
	\includegraphics[width=1\linewidth]{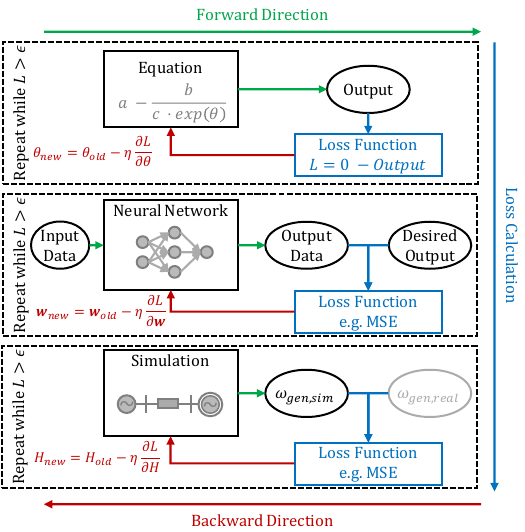}
	\caption{Comparison of the optimization process for the function, ANNs and the power system simulation}
	\label{fig:trainingprocess}
\end{figure}

Exactly the same process can be applied to power system simulations as shown in Fig. \ref{fig:trainingprocess}. More concrete, the process consists of the following steps:

\begin{enumerate}
\item Run the simulation. During the simulation, the AD tool automatically builds a computational graph.
\item Use the output of the simulation to calculate a loss, e.g. the MSE loss between output of the simulation and desired output.
\item Use the AD tool to calculate the gradients of the loss function with respect to the simulation parameters by traversing the computational graph in the backward direction.
\item Use the gradients to calculate new values of the simulation parameters by applying the gradient descent method described in (\ref{eq:graddesc}).
\item Repeat steps 1) - 4) until the loss is below a predetermined threshold.

\end{enumerate}

In the example of the SMIB system, $H$ can be adapted, so that the time series data of $\Delta \omega_{gen,sim}$ matches the real data. This process can also be applied to multiple parameters simultaneously. Additionally, it is not only applicable for time series matching, but for many use cases. The prerequisite is, that the desired behavior can be described by a loss function. This is true for many examples such as determining controller parameters for optimal damping, converting blackbox models to interpretable models, optimal power flow problems and more.

\section{Implementation Details}
To enable a proof of concept for the new method, we implemented a power system simulation in Python. The implementation is based on the works of Haugdal and Uhlen \cite{Haugdal.08Jan21}, but a re-implementation and simplification was necessary to allow the gradient tracking. All optimizable parameters are implemented as PyTorch tensors. This enables the PyTorch backend to serve as an AD tool and automatically generate the computational graphs. Two challenges that arise with the approach can be solved in the implementation. The first challenge is known from deep neural networks as the vanishing or exploding gradient problem. The other issue is, that the gradient descent will get stuck in local optima and is therefore highly dependent on the choice of initial parameters.

\subsection{Vanishing and Exploding Gradients}
The long chain of multiplications when applying the chainrule of differentiation can in theory lead to the total gradient to vanish or explode. As an example, assume the chain in (\ref{eq:chainrule}) consists of $1000$ factors, and every local gradient has a value of $0.9$. In this case, the total gradient would vanish to $1.7e-46$. Similarly, the gradient can explode for values greater than one. Fortunately, the utilization of normalizing the output and using the per unit system, reduces the problem. We found that the gradients of the loss function with respect to certain parameters are remarkably stable during an optimization process. When adapting the learning rate accordingly, we didn't find any issue with exploding or vanishing gradients when the simulation itself was stable.

The gradients do tend to explode though in cases where the simulation itself is unstable. While the gradients mostly point in the right direction, and therefore lead the simulation towards a more stable operation, an instability in the simulation can lead to an instable optimization process, because the large gradients adapt the parameters too fast. To solve this issue, we employ a custom optimizer with a predefined maximum step size that limits the parameter adaption per optimization step. This stabilizes the optimization process, which means that initial values can be used for parameters that lead to unstable simulations.

\subsection{Local Optima}
\begin{figure}
	\centering
	\includegraphics[width=1\linewidth]{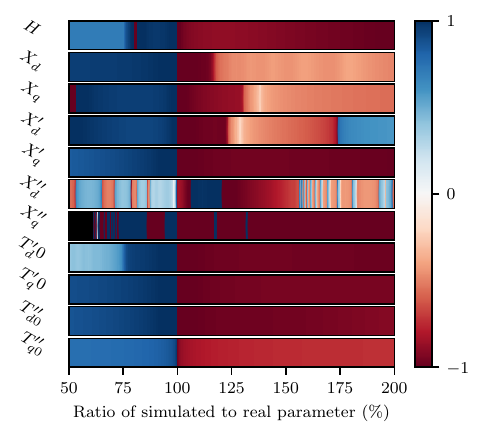}
	\caption{Gradients of the loss function with respect to SMIB parameters}
	\label{fig:gradients}
\end{figure}

The second challenge is the existence of local optima. While gradient descent optimizations exist that use a momentum function in order to overcome those, local optima pose an inherent problem to the presented method. Fortunately, in the experiments we ran so far, local optima were an uncommon occurrence. This is shown in Fig. \ref{fig:gradients}. In order to find local optima, we varied the parameters of the SMIB system between $50 \%$ and $200 \%$ of their original value and calculated the MSE-loss between the original and new trajectory of $\Delta \omega$. The corresponding normalized gradients of the loss function with respect to the parameters are shown in Fig. \ref{fig:gradients}. For an ideally convex optimization, the gradients would be seen as blue, positive values if the parameter was reduced, and red, negative if it was increased. This is true for many parameters, but it can be seen, that especially $X^{\prime\prime}_d$, and $X^{\prime\prime}_q$ do have a number of local optima. Therefore, the initial guess of those parameters is quite relevant.

As it can be quite time consuming to search good initial parameter guesses, we aimed to tackle this issue in the implementation. By using PyTorch tensors for the power system simulation, the implementation is inherently vectorized. Therefore, it is computationally inexpensive to run the simulation process with multiple values for each parameter in parallel. Due to this implementation, a vector of initial guesses can be used for all optimizable parameters. Using a vector of initial guesses significantly decreases the probability that the trajectories from all starting points get stuck in local optima and therefore increases the probability of finding a global optimum. Therefore, the utilization of multiple initial guesses makes the search for good initial guesses obsolete.

\section{Results}
\label{sec:results}
To explore the effectiveness of the proposed approach, we show two use-cases. The first one is the already mentioned parameter estimation for the SMIB system, the second experiment aims to tune the parameters of a power system stabilizer (PSS) in order to facilitate maximum damping. The spirit behind the experiments is to maintain simplicity, in order to show the results in a straightforward manner. The code to replicate those experiments is available on GitHub \cite{Kordowich.2023}.

\subsection{Parameter Identification of a SMIB system}
\subsubsection{Model}
For the first experiment, we pretend the inertia constant $H_{gen}$ of the SMIB system is unknown. The SMIB system consists of two 6th order generator models, of which one is very large and can therefore be considered an infinite busbar. The parameters are taken from Kundur \cite{Kundur.2007} and are equivalent to the model used in \cite{Haugdal.08Jan21}. No governor, exciter or power system stabilizer are used for the first experiment.
\subsubsection{Optimization Process}
To apply the parameter identification process, "real data" which the simulation can be aligned with are necessary. To create those, we ran a simulation of the SMIB system with an inertia constant $H_{gen, real}$ of \qty{3.5}{s} in DIgSILENT's PowerFactory. During the simulation, a short circuit is simulated at $t = \qty{1}{s}$ until $t = \qty{1.05}{s}$. This induces an oscillation in the trajectory of $\Delta \omega$ which is then exported. Afterwards, the process described in subsection \ref{sec:bp} can be used to fit the simulation parameter $H_{gen, sim}$ to the "real world" parameter $H_{gen, real}$. For this purpose, we simulate the same SMIB system disturbed by a short circuit using the power system simulation implemented in Python in order to track the gradients. The initial guess for the parameter $H_{gen, sim}$ is \qty{8}{\second}. As a loss function we use the MSE between the trajectory of $\Delta\omega_{gen,real}$ and $\Delta\omega_{gen,sim}$ given in (\ref{eq:loss}).

\begin{figure}
	\centering
	\includegraphics[width=1\linewidth]{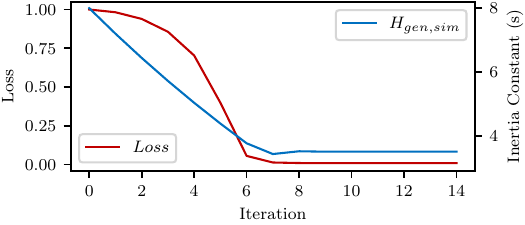}
	\caption{Loss and inertia constant during the optimization process}
	\label{fig:lossplot}
\end{figure}

\begin{figure}
	\centering
	\includegraphics[width=1\linewidth]{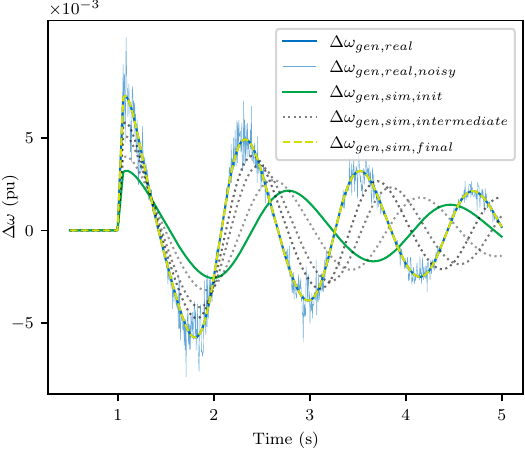}
	\caption{$\Delta \omega$ trajectories of real-world and simulated generator during the optimization process}
	\label{fig:omegahopt}
\end{figure}

The optimization process iteratively adapts the parameter, which decreases the normalized loss as shown in Fig. \ref{fig:lossplot}. Fig.~\ref{fig:omegahopt} illustrates how the trajectory of $\Delta\omega_{gen,sim}$ progressively converges toward the trajectory of the real-world scenario, eventually achieving a perfect match. The optimization stops when $H_{gen, sim}$ changes less than $1e-6$ within one optimization step. The final value is $H_{gen, sim} = 3.5001$ which is equivalent of an error of \qty{0.003}{\percent}.


\subsubsection{Influence of Noise}
In order to test the influence of noise on the optimization, we added gaussian noise to the signal as shown in Fig. \ref{fig:omegahopt} as a thin blue line. For different noise levels, the error slightly increased to up to \qty{0.2}{\percent} as shown in Tab. \ref{tab:noise}.
\begin{table}[h]
	\centering
	\renewcommand{\arraystretch}{1.1}
	\caption{Parameters of the SMIB system}
	\label{tab:noise}
	\begin{tabular}{|l||c|c|c|c|}
		\hline
		Noise Level & \qty{0}{\%}  & \qty{5}{\%} & \qty{10}{\%} & \qty{20}{\%} \\
		\hline
		$H_{gen, sim, final}$ & \qty{3.5001}{\second} & \qty{3.5011}{\second} & \qty{3.4994}{\second} & \qty{3.5075}{\second} \\
		\hline
		Error & \qty{0.003}{\%} & \qty{0.03}{\%}& \qty{0.02}{\%}& \qty{0.2}{\%} \\
		\hline
	\end{tabular}
\end{table}
Even though the error slightly increases, the optimization is extremely robust to noise, as the absolute error is still very low. The reason for this is, that the loss function takes the whole trajectory into account, and therefore inherently averages out noise.

\subsection{Parameter Optimisation of a Power System Stabilizer}
\begin{figure}
	\centering
	\includegraphics[width=1\linewidth, trim={0 30 0 0},clip]{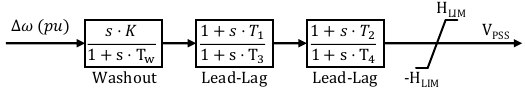}
	\caption{The power system stabilizer model STAB1}
	\label{fig:STAB1}
\end{figure}

The goal of the second experiment is to tune the parameters of a PSS in order to facilitate maximum damping. This is a classic parameter optimization problem. For this purpose we extended the previously mentioned model with an automatic voltage regulation system (AVR) and a power system stabilizer. In the spirit of simplicity for the experiments, we chose the simple exciter (SEXS) model as an AVR and the STAB1 model shown in Fig. \ref{fig:STAB1} as a PSS \cite{NEPLANAG.}.
\begin{figure}
	\centering
	\includegraphics[width=1\linewidth]{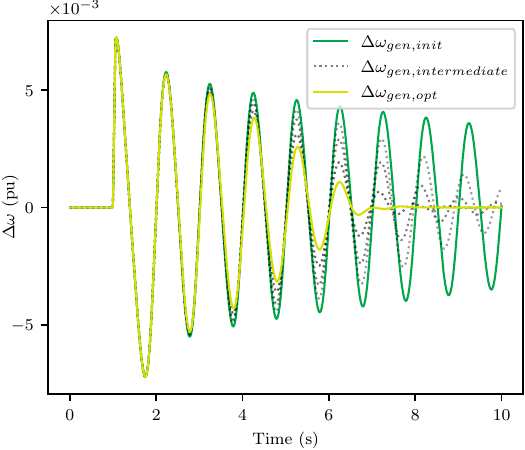}
	\caption{$\Delta \omega$ trajectories of the simulated generator during the optimization of PSS parameters}
	\label{fig:omegapssopt}
\end{figure}
\begin{table}
	\centering
	\renewcommand{\arraystretch}{1.1}
	\caption{Parameters of the PSS}
	\setlength\tabcolsep{4pt}
	\label{tab:pss}
	\begin{tabular}{|c||c|c|c|c|c|c|c|}
		\hline
		Parameter & $K$ & $T_{w}$ & $T_1$ & $T_3$ & $T_2$ & $T_4$ & $H_{lim}$ \\
		\hline
		Initial Guess &$40$&\qty{11}{\second}& \qty{0.08}{\second} & \qty{0.10}{\second} & \qty{0.50}{\second} & \qty{0.05}{\second} & \qty{0.03}{pu} \\
		\hline
		\makecell{Optimized \\ Value} &$56$&\qty{3.6}{\second}& \qty{0.12}{\second} & \qty{0.04}{\second} & \qty{0.73}{\second} & \qty{0.02}{\second} & \qty{0.03}{pu} \\
		\hline
	\end{tabular}
\end{table}
The initial guesses of the relevant parameters of the PSS are listed in Tab. \ref{tab:pss}. All those parameters are tuned simultaneously by calculating all the respective gradients at once using the AD tool. For the optimization process, a suitable loss function must be selected. As the maximum peak is mostly determined by the short circuit, and can not be influenced by the PSS, we chose the mean absolute error as a loss function, as it is more robust to outliers than the MSE:
\begin{equation}
	L(K, T_w, T_1, T_2, T_3, T_4) =\frac{1}{N} \sum_{t=\qty{1}{\second}}^{t=\qty{10}{\second}}\Delta\omega_{sim, t}
	\label{eq:MAE}
\end{equation}
The optimization goal is therefore to reduce the oscillation of $\Delta\omega_{sim, t}$ to zero as fast as possible. In Fig. \ref{fig:omegapssopt} it is clear, that the oscillation decays only very slowly when using the suboptimal values of the initial guess. During the optimization, once again the loss decreases, and the oscillation therefore decays faster and faster. This is achieved by optimizing the parameters towards the final values given in Tab. \ref{tab:pss}. Note that the parameter $H_{lim}$ for the voltage limiting is not optimized. As no optimal parameters are known for this problem, no error can be given for this experiment. It can be seen though, that the parameter optimization is successful as the oscillation is damped significantly better than with the initial guess.

\section{Discussion}
The experiments in the previous section show promising results. They imply that the tool has a generic applicability for a wide range of problems. An advantage is the simplicity of the approach and the easy implementation of the optimization itself. However, it's important to acknowledge that there are several uncertainties that need yet to be investigated. 

Even though we did not face such a problem so far, the process is not suited for strongly multimodal optimization problems that have many local optima. Even though the challenge of local optima can be extenuated by using a vector of initial guesses, the probability to get stuck in a local optimum increases with their quantity. As of now, it is unclear how many real world power system optimization problems are too multimodal for the optimization.

Additionally, while it is easy to implement the optimization itself, it is quite time consuming to implement and test the power system simulation itself in Python. An implementation using a modular concept, as demonstrated in \cite{Haugdal.08Jan21}, is feasible. Nevertheless, creating a generic power system simulation library is a substantial undertaking.

While the proposed optimization has some benefits over applying PINNs instead, it must be noted, that the process is computationally expensive in comparison. Python is inherently relatively slow, and the tracking and backpropagation of the gradients does take a certain amount of time. This has not been an issue for the simple experiments so far, but the scalability of the approach must be investigated.

The results show, that the optimization process can correctly identify multiple parameters simultaneously from a single trajectory, and is therefore essentially a one-shot learning approach. It must be noted though, that the optimization only finds "a", not necessarily "the" optimal solution. For larger systems, it is conceivable, that multiple sets of parameters can minimize the loss function equally well. In this case, the one-shot learning property is lost, and more additional information is necessary.

\section{Conclusion}
This paper proposes a gradient descent based optimization approach for power system simulations, that relies on tracking the gradients during a simulation using an AD tool, namely PyTorch. The paper shows the theoretical foundation, practical implementation and first results of use-cases. In the experiments, we show that the optimization process can identify uncertain power system parameters from a single trajectory of a dynamic process, and optimize controller parameters with respect to the goal of power swing damping. The results imply that the optimization process is generically applicable and has the potential to help solve a wide range of optimization problems. Additionally, it significantly reduces the number of simulations necessary for the optimization of multiple parameters. Future works will demonstrate more use cases and examine the scalability of the approach.
\newpage

\bibliographystyle{IEEEtran}
\bibliography{IEEEabrv,literatur_bitex}

\end{document}